\documentstyle[12pt,titlepage]{article}
\title{Monopole Percolation and The Universality Class of the Chiral
Transition in Four Flavor Noncompact Lattice QED}

\author{by\\[8pt]  A. Kocic\\ Theory Division\\CERN\\ CH-1211 Geneva
23, Switzerland\and J. B. Kogut\\ Physics Department\\ 1110 West Green
Street\\University of Illinois\\ Urbana, IL  61801-3080\and
 K. C. Wang\\ Australian National University\\ Box 4, G.P.O.\\
Carberra, A.C.T. 2600 Australia}

\begin{document}

\def\bra{\langle} \def\ket{\rangle}

\maketitle

\begin{abstract}
We simulate four flavor noncompact lattice QED using
the Hybrid Monte Carlo algorithm on $10^4$ and $16^4$ lattices.
Measurements of the monopole susceptibility and the percolation order
parameter indicate a transition at $\beta = {1/e^2} = .205(5)$ with
critical behavior in the universality class of four dimensional
percolation.  We present accurate chiral condensate measurements and
monitor finite size effects carefully. The chiral condensate data
supports the existence of a power-law transition at $\beta = .205$ in
the same universality class as the chiral transition in the two flavor
model.  The resulting equation of state predicts the mass ratio
$m_\pi^2/m_\sigma^2$ in good agreement with spectrum calculations while
the hypothesis of a logarithmically improved mean field theory fails
qualitatively.

\end{abstract}

\section{Introduction}
Computer simulations of noncompact lattice QED
have proven to be very intriguing and surprising.  For two and four
flavors of light dynamical fermions a continuous chiral symmetry
breaking transition was found some time ago$^{[1]}$.  Recent large
scale simulations of the $N_f = 2$ flavor model have given results
consistent with a non-mean field transition which might signal the
existence of a non-trivial fixed point$^{[2]}$.  It was also found that
the chiral transition is essentially coincident with a monopole
condensation transition whose critical properties are in the
universality class of four dimensional percolation$^{[3]}$.  Since
conventional physical pictures of chiral symmetry breaking$^{[4]}$
indicate that monopole condensation implies chiral symmetry breaking,
these results suggest that theories of fundamental charges and
monopoles provide a natural scenerio for nontrivial ultraviolet
behavior.  The purpose of the present article is to investigate whether
these results extend to the $N_f = 4$ theory where fermion screening is
even more severe.  We shall provide evidence for the remarkably simple
and intriguing result that the $N_f = 2$ and 4 theories lie in the same
universality class.  Although fermion screening will shift the theory's
critical point to stronger coupling than seen in the $N_f = 2$ theory,
the critical indices governing both monopole condensation and chiral
symmetry breaking will be essentially identical.  These results
underscore the fact that renormalization group studies of noncompact
lattice QED which are based on  perturbative ideas borrowed from weakly
coupled continuum QED are {\em not} justified here.  Theories with
fundamental monopoles are intrinsically nonperturbative  because of
the Dirac quantization condition.  The numerical evidence that we
shall present supporting the claim that the $N_f=2$ and 4 theories
share the same critical behavior is inexplicable perturbatively.

We begin by summarizing the main points of the later sections of this
paper.  The reader is referred to previous simulations and analytic
papers for additional background.  In Sec.~2 of this paper we consider
the evidence for a monopole condensation transition in the $N_f=4$
theory.  Finite size scaling analyses of our $10^4$ and $16^4$ data
produce estimates for the critical indices of this transition which
place it in the universality of four dimensional percolation.  In
Sec.~3 we consider our $10^4$ and $16^4$ measurements of the chiral
condensate.  The $10^4$ measurements are particularly extensive and
involved couplings $\beta={1/e^2}=.22$, .215, .21, .205, .20, .195,
.19, .185 and bare fermion masses of $m=.005$, .01, .02, .03, .04, .05,
.06 and .07.  Comparing these results with past measurements and our
$16^4$ measurements at $m=.03$, we conclude that finite size effects
are sufficiently small only for $m\geq.03$ on the $10^4$ lattice to
permit quantitative analysis.  The $10^4$ data at $m\geq.03$ are
consistent with a power-law chiral transition between $\beta=.210$ and
.205 with a critical index $\delta=2.3(1)$.  The chiral condensate data
falls on a universal Equation of State with the critical point
$\beta_c=.205$ and the critical indixes $\delta=2.31$ and
$\beta_{mag}=.764$ which deviate significantly from mean field values.
In Sec.~4 we consider the spectroscopy data of Ref. [5] and
concentrate on the ratio $R(\beta,m)=m_\pi^2/m_\sigma^2$ since it can
be predicted from the model's Equation of State with no free
parameters$^{[6]}$.  We show that the Equation of State of Sec.~3
accounts for the data well while the Equation of State of Ref.~[5]
which is based on a logarithmically trivial model fails qualitatively.
Finally, in Sec.~5 we observe that the Equation of State found in
Sec.~2 for the four flavor model is essentially identical to that of
the two flavor model studied for different parameters, on different
lattice sizes by different algorithms on different computers.  This
result indicates that the two models are in the same universality
class.  In Sec.~6 additional work is proposed to challenge the simple
physical picture that the results of Sec.~5 suggest.

\section{Monopole Condensation in the Four\protect\\ Flavor Model} We
have discussed monopole percolation in quenched noncompact lattice QED
in Ref. [2] and in the two flavor model in Ref. [3].  The quenched
model was analyzed with considerable accuracy because photon
configurations could be generated using FFT methods.  Thousands of
statistically independent configurations could be generated and
analyzed on lattices ranging up to $24^4$.  Finite size scaling
analyses then produced critical exponents of the monopole percolation
transition at coupling $\beta=.244$ in excellent agreement with four
dimensional percolation.  The same measurements on two flavor
configurations on $8^4-18^4$ lattices produced the same set of critical
indices, although the statistical uncertainties were considerably
larger due to the relative slowness of the hybrid molecular dynamics
algorithm.  Fermion feedback shifted the monopole condensation
transition to stronger coupling, $\beta=.225(5)$.  In the four flavor
model the hybrid molecular dynamics algorithm was replaced by the
hybrid monte carlo scheme which is free of systematic errors.  It
proved easy to keep the time step of the algorithm reasonably large
while maintaining a high (50\%-80\%) acceptance rate so that the
algorithm explored phase space relatively efficiently.  The time step
$dt$ was tuned proportional to the bare fermion mass $m$:  for $m =
.005$, $dt = .005$; for $m = .01$, $dt = .01$; for $m\geq .02$, $dt =
.02$.  Measurements were taken after unit time intervals, and between
250--500 time intervals were generated at each $m$ and $\beta$.  In
fact, at crucial couplings such as $\beta = .205$ we doubled our
statistics.  In that case we required that the product of the
acceptance rate times the number of trial trajectories (a trajectory is
defined to be a unit time interval, or $(dt)^{-1}$ sweeps) be at least
500.  Statistical errors in the observables were then estimated by
standard binning methods and those statistical errors are recorded in
our tables below of raw data.

As in past studies we measured the monopole percolation order parameter
$M = n_{max}/n_{tot}$, its susceptibility $\chi$ and the monopole
concentration $\bra\ell\ket/N_{\ell}{}^{[7]}$.  Recall that $n_{max}$ is
the number of links with nonvanishing monopole current in the largest
connected monopole cluster on the lattice and $n_{tot}$ is the total
number of such links.  The susceptibility $\chi$ is essentially the
variance in the order parameter $M$.  The monopole concentration
$\bra\ell\ket/N_{\ell}$ is the average amount of monopole current on
the lattice per link.  These observables and bond percolation in
general have been discussed more fully in Ref.~[7] and~[3], and the
reader should consult those discussions for additional theoretical
background.  The monopole susceptibility, order parameter and
concentration measurements on our $10^4$ lattices are recorded in
Tables 1--3.  In Table 4 we list our $16^4$ data which we collected at
only one fermion mass value $m = .03$ due to our limited computer
resources.  Consider the monopole susceptibility first.  In Fig. 1 we
plot the $10^4$ and $16^4$ data vs.\ $\beta={1/e}^2$ at $m = .03$.
Clearly the peak is considerably higher and sharper on the $16^4$
lattice.  We can estimate the ratio of the susceptibility to the
correlation length critical indices for monopole condensation,
$\gamma_{mon}/\nu_{mon}$, from finite size scaling which states that
the maximum of $\chi$ on a $L^4$ lattice should grow as,
\begin{equation}
\chi_{max}\sim L^{\gamma_{mon}/\nu_{mon}} \label{2.1}
\end{equation}
 From the tables on Fig. 1 we calculate,
\begin{equation}
\gamma_{mon}/\nu_{mon} = 2.29(9)
\end{equation}
which should be compared to the value 2.25(1) found in the quenched
model$^{[3]}$.  The peak in $\chi$ occurs at $\beta=.210$ in the four
flavor model as compared to $\beta=.244$ in the quenched model.  So
fermion, screening has shifted the monopole percolation transition to
stronger coupling but apparently has not affected it otherwise.  The
scaling behavior of the order parameter at the critical point also
yields another ratio of critical indices,
\begin{equation}
M(\beta=.210,L)\sim L^{-\beta_{mon}/{\nu}_{mon}}
\end{equation}
where $\beta_{mon}$ is the ``magnetic'' exponent which governs the
$\beta$ dependence of the order parameter in the condensed phase in the
thermodynamic limit,
\begin{equation}
M(\beta) = C(\beta_c-\beta)^{\beta_{mon}},\qquad
\beta<\beta_c\cong.210
\end{equation}
Since $M = .239(9)$ on the $10^4$ lattice at $\beta =
.210$ and $M = .1452(41)$ on the $16^4$ lattice at $\beta = .210$, Eq.
(3) implies that
\begin{equation}
\beta_{mon}/\nu_{mon} = 0.9(1)
\end{equation}
This result should be compared to the quenched model's value of .88(2)
and the supposedly exact value for four dimensional percolation of
.875.  Again, fermion screening appears to have left the critical
behavior of the monopole percolation transition unchanged.

Finally, in Fig. 2 we plot the order parameter $M$ and the monopole
concentration $\bra\ell\ket/N_{\ell}$ on the $16^4$ lattice.  Note that
the monopole concentration varies smoothly between a value of .12 at
$\beta = .22$ and .17 at $\beta = .19$ while the order parameter turns
on abruptly at $\beta_c = .205(5)$.  Following the discussion of bond
percolation in Ref. [7], recall that the critical concentration of
occupied bonds that induces the percolation transition in four
dimensions is .16(1).  Since $\bra\ell\ket/N_{\ell}$ varies smoothly as
a function of $\beta$, one could parametrize monopole condensation as a
function of $\bra\ell\ket/N_{\ell}$ and change the language of the
transition to resemble that of bond percolation in greater detail.
Since Eq. (2) and (5) are compatible with the critical
indices of the traditional percolation transition, the parallel between
the two transitions seems very reasonable.

In the quenched and two flavor model much more extensive measurements
of the monopole transition were made, so the scaling laws
Eq. (1) and (3) were checked in greater detail$^{[2]}$.  It
would be worthwhile to do the same for the four flavor model as well,
given more computer resources.

A final comment on the bare fermion mass dependence of the monopole
data.  We note from Table 1 that as $m$ increases from .005 to .07, the
peak in $\chi$ shifts from .205 to .215.  This is the expected
trend---as $m$ increases fermion screening is suppressed and the
monopole transition moves toward its quenched coupling of .244.  Note,
however, that as $m$ varies the height of the peak in $\chi$ does not
change significantly.

\section{Chiral Condensate Measurement and the Equation of State}

Now consider the chiral transition.  We accumulated extensive data on
the average plaquette and the chiral condensate
${\bra\overline{\Psi}\Psi\ket}$ recorded in Tables 5 and 6
respectively.  The relatively small error bars were deduced from
standard binning procedures.  As we have seen in $N_f=0^{[8]}$ and
2$^{[2]}$ studies, accuracy is essential in searching for the critical
behavior in noncompact lattice QED ${\bra\overline{\Psi}\Psi\ket}$
measurements.  It is also necessary to have data over a wide range in
$m$ and $\beta$, to meaningfully distinguish between different
theoretical models.  Since $10^4$ is not a particularly large lattice,
we must study finite size effects carefully.  Recall that the finite
size effects in the quenched ${\bra\overline{\Psi}\Psi\ket}$ data were
very small and bare fermion masses in the range .001--.005 could be
used to search for critical behavior$^{[8]}$.  For $N_f=2$ the finite
size effects were larger and ${\bra\overline{\Psi}\Psi\ket}$ data at
$m=.005$ were sightly suppressed on $10^4$ lattices as compared to
$16^4{}^{[2]}$.  Data at $m\geq.01$ did not display statistically
significant finite size effects.  The finite size effects for the
$N_f=4$ model are even larger.  This is illustrated in Table 7 where we
compare a sample of our $10^4$ data with the $12^4$ data of Ref. [5].
The $m=.04$ data show no size dependence while the $m=.02$ and $.01$
${\bra\overline{\Psi}\Psi\ket}$ measurements are suppressed on the
smaller lattice.  Since the lattice has only been scaled by 20\% in
linear dimensions between $10^4$ and $12^4$, these finite size effects
in ${\bra\bar{\Psi}\Psi\ket}$ are very dangerous.  Therefore, we must
discard our $10^4$ data at $m=.005$, $.01$ and $.02$ when searching for
the bulk critical behavior in $\bra\overline{\Psi}\Psi\ket$.  This
finding convinced us to accumulate data out to $m=.07$.  We did not
proceed further because once $m$ becomes a fair fraction of unity,
non-universal lattice artifacts will plague the values of
${\bra\overline{\Psi}\Psi\ket}$.  The reader can check that the $m=.03$
${\bra\overline{\Psi}\Psi\ket}$ measurements of Table 6 on a $10^4$
lattice and the $m=.03$ ${\bra\overline{\Psi}\Psi\ket}$ measurements of
Table 4 are in good agreement.

In order to argue that there is a chiral transition hiding in the data
of Table 6, we must make a hypothesis concerning the character of the
critical point.  The hypothesis we have been pursuing in recent work is
that the critical behavior is controlled by a power-law divergent
correlation length  ${\xi\sim|\beta-\beta_c|^{-\nu}}$ in the chiral
$m\rightarrow0$ limit.  This is the simplest nontrivial hypothesis we
can make and since it occurs many times in statistical mechanics,
effective data analysis strategies are known and well-understood.  In
particular, there should be a scaling region around the critical point
where ${\bra\overline{\Psi}\Psi\ket}$ satisfies a universal equation of
state,
\begin{equation}
\frac{m}{\bra\overline{\Psi}\Psi\ket^\delta}=
f\left(\frac{\Delta\beta}{\bra\overline{\Psi}\Psi\ket^{1/\beta_{mag}}}
\right)
\end{equation}
where $\Delta\beta=\beta-\beta_c$ and the parameters
$\delta$ and $\beta_{mag}$ are familiar critical indices.  In
particular, at $\beta=\beta_c$, $\delta$ controls the response of the
chiral condensate to the symmetry breaking field $m$,
\begin{equation}
\bra\bar{\Psi}\Psi\ket\sim m^{1/\delta}\qquad(\beta=\beta_c)
\end{equation}
The index $\beta_{mag}$ controls the shape of
$\bra\overline{\Psi}\Psi\ket$ as a function of coupling $\beta={1/e^2}$
within the broken-symmetry phase,
\begin{equation}
\bra\bar{\Psi}\Psi\ket=
D(\beta-\beta_c)^{\beta_{mag}}\qquad(\beta<\beta_c)
\end{equation}
in the chiral limit $m\rightarrow0$.  In order to find
$\beta_c$ and measure $\delta$ we consider Eq. (7) and plot
$\ln\bra\bar{\Psi}\Psi\ket$ vs.\ $\ln(m)$ from Table 6.  The
fixed-$\beta$ plots are shown in Fig.\ 3.  The coupling $\beta_c=.205$
fits a power-law Eq. (7) excellently with a value of $\delta=2.31$.
Since the data at other $\beta$ values do not deviate far from straight
lines,  we see again the need for accurate $\bra\bar{\Psi}\Psi\ket$
data to proceed.  Another slightly different and better way to find
$\delta$ and $\beta_c$ is to plot $-1/\ln(m)$ vs.\
$-1/\ln\bra\bar{\Psi}\Psi\ket$ at each coupling.  At the critical
coupling such a curve should be linear with a slope of $\delta$ {\em
and} it should pass through the origin.  The data of Table 6 for
$m=.03$--.07 are plotted in this fashion in Fig.\ 4 and only
$\beta_c=.21$ or $.205$ emerge as candidates for the critical point.
The straight lines in the figure give $\delta=2.2(1)$ in good
agreement, not surprisingly, with Fig. 3.  The reader should note that
the data for other $\beta$ values do not lie on straight lines which
pass through the origin in Fig.\ 4.

Two comments are in order at this point.  First, if the analysis of
Figs.~3--4 is correct, then monopole condensation and chiral symmetry
breaking would be coincident transitions.  This is very satisfying and
would fit into the physical picture of Ref. [4] nicely.  The estimate
of $\beta_c=.205-.210$ agrees with our earlier, cruder work and also
with the interesting and different approach of Ref. [9].  It disagrees
with the chiral critical point of Ref. [5] and we shall discuss this
point in detail below.  The second comment consists of the observation
that $\delta=2.31$ was exactly the critical exponent found by similar
procedures in a much larger scale simulation of the two flavor
model$^{[2]}$.  This result, which may be a crucial key into the physics
of noncompact lattice QED, will be discussed in Sec. 5 below.

In order to investigate the Equation of State Eq. (6) we need an
estimate of $\beta_{mag}$ in addition to $\beta_c$ and $\delta$.
Recall that $\beta_{mag}$ is related to $\gamma$, the ``magnetic''
susceptibility index and $\delta$ through the relation
$\beta_{mag}=\gamma$/($\delta-1$).  We first investigate the hypothesis
$\gamma=1$ so that $\beta_{mag}=.764$.  The hypothesis $\gamma=1$ will
prove to be self-consistent.  It is a reasonable choice because it is
true in the quenched model, as seen in simulations$^{[8]}$ and analytic
Schwinger-Dyson studies$^{[10]}$, and it is also true in mean field
theory.  Regardless of the motivation for $\beta_{mag}=.764$, Fig. 5 of
Eq. (6) follows.  The data appear to lie around a straight line which
is approximately,
\begin{equation}
f(x)=-5.3125 x +1.15
\end{equation}
As discussed in Ref.~[6] a linear universal function $f$ is only
possible if $\gamma$ is precisely unity, so the self-consistency of
this discussion is certainly pleasing.

If Fig. 5 and Eq. (9) are really true, then the chiral transition is
coincident with monopole condensation and its critical indices deviate
significantly from mean field theory (where $\delta=3$,
$\beta_{mag}={1/2}$).  However, since Fig. 5 can be viewed as a ``fit''
depending on several parameters ($\beta_c$, $\delta$ and $\beta_{mag}$)
it's significance and uniqueness are open to argument.  Other
hypotheses for the character of the phase transition, complete with
their parameters, might describe the data just as well or even better.
It is important to confront the Equation of State fit to other
independent challenges.

\section{Spectroscopy and the Equation of State}

There is much more content in the Equation of State and the universal
function of $f$ than the chiral condensate.  In Ref.~[6] we develop
the scaling theory of mass ratios and apply it to chiral transitions.
In particular, the ratio of the squares of the pion and sigma masses,
\begin{equation}
R(\beta,m)={m^2_\pi\over m^2_\sigma}
\end{equation}
is particularly
illuminating because it is a dimensionless renormalization group
invariant quantity and because it is highly constrained by chiral
symmetry.  In fact, as discussed in Ref.~[6] it is completely
determined within the scaling region by the universal function
$f^{[6]}$,  %
\begin{equation}
R = {m^2_\pi\over m^2_\sigma} =
\left(\delta-{x\over\beta_{mag}}{f'(x)\over f(x)}\right)^{-1}
\end{equation}
The general shapes of the curves of $R(\beta,m)$ at fixed couplings are
informative and easily understood.  At the critical point $x=0$, so $R$
reduces to
\begin{equation}
R(\beta_c,m)={1\over\delta}
\end{equation}
So, the curve should be flat at $\beta_c$ and give another estimate of
$\delta$.  For ${\beta<\beta_c}$ in the broken symmetry phase, the
curves should be concave downward and intersect the origin because the
pion is massless in the $m\to 0$ limit.  For $\beta<\beta_c$ in
the symmetric phase, the curves should be concave upward and intersect
$R=1$ at $m=0$ because the pion and the sigma should become partners of
a chiral multiplet.

In Ref.~[2] we showed that $R$, as determined analytically from that
theory's Equation of State, fit the two flavor model's spectroscopy
data very well and had the general features expected from chiral
symmetry consideration.   To do the same exercise for the four flavor
model we borrow the $m_\pi$ and $m_\sigma$ spectroscopy data from
Ref.~[5].  Tables of pion and sigma masses can be found there for
$\beta=.22$, .21, .20, .19, .18, .17, and $m=.16$, .09, .04, .02 and
$.01$.  We accept this data and plot the ratio $R(\beta,m)$ vs.~$M$
at fixed $\beta$ values in Fig. 6 with the error bars as given in
Ref.~[5].  We note that $R$ is flat for $\beta=.20-.21$ with a value
near $.4$ implying $\delta\approx2.5$.  In Fig.~7 we plot Eq. (11)
with the universal function Eq. (9) choosing the same couplings
$\beta=.22$--.17 as the spectoscopy data of Fig~6.  The general
agreement between the plots is very satisfying, although the
theoretical plot lies slightly higher than the data.

We close with a
remark about the meson spectrum calculation and the ratio $R$.  The
computer calculation of $m_\pi^2$ and $m_\sigma^2$ are done in the
``valence quark'' picture which ignores possible mixing with pure
multi-photon states.  That mixing is proportional to the overlap of
the lowest lying two-fermion (positronium) and photonium (glueball)
states.  If the lowest lying photonium state is relatively heavy, as
expected, the mixing would be small.  The success of our calculation
Fig.~7 certainly suggests this, but an explicit verification would be
best.

\section{The Failure of Logarithmically Improved Mean Field Theory}

As discussed in Sec.~3, the chiral condensate data may be fit by very
different hypotheses.  In particular, in Ref. [5] a logarithmically
improved $O(2)$ sigma model is used for fitting purposes and it fits
the data very well.  The Equation of State reads,
\begin{equation}
m=2\sigma V'_{eff}(\sigma^2)=
\tau{\sigma\over \ell n^p|\sigma^{-1}|}+
\theta{\sigma^3\over \ell n|\sigma^{-1}|}
\end{equation}
where $\sigma=\bra\bar\psi\psi\ket$,
$\tau=\tau_1\theta(1-\beta/\beta_c)$ and $\theta^{-1}
=\theta_o+\theta_1(1-\beta/\beta_c)$.  Choosing specific values for the
five parameters $\beta_c$, $p$, $\tau_1$, $\theta_c$ and $\theta_1$, a
very good fit to the data is found.  The resulting chiral transition
occurs at $\beta_c=.186(1)$ with mean field critical indices built in.
Two comments are in order.  First, since the fit involves five
parameters, its significance is certainly debatable.  However, our
new $\bra\bar\psi\psi\ket$ data for $\beta=.22$--.185 and $m=.03$--.07
satisfies Eq. (13) very well as shown in Fig. 8.  This is certainly
expected since our range of $\beta$ and $m$ values lies within those
considered in the original fit of Ref.~[5].  Second,
$\beta_c=.186(1)$ lies deep within the monopole condensate phase found
here.  We would expect that all the physics is pushed to the cutoff at
such a strong coupling and that the size of bound states, etc. would be
on the order of the lattice spacing for $\beta\approx.186$.

Let us subject Eq. (13) to the same test that Eq. (6), (9) and
(11) have just passed.  In particular, from Sec.~8 of Ref.~[5]
we read off a formula for $R=m^2_\pi/m^2_\sigma$ calculated in the
logarithmically-improved $O(2)$ sigma model,
\addtocounter{equation}{1}
$$
R={m^2_\pi\over m^2_\sigma}=
\left(1+2\sigma^2V''_{eff}(\sigma^2)/V'_{eff}(\sigma^2)\right)^{-1}
\eqno(\theequation \mbox{a})
$$
and
$$
V''_{eff}(\sigma^2)=
{p\tau\over 4\sigma^2\ln^{p+1}|\sigma^{-1}|}+
{\theta\over 2\ln|\sigma^{-1}|}+{\theta\over 4\ln^2|\sigma^{-1}|}
\eqno(\theequation \mbox{b}) \bigskip
$$
The important point about Eq. (14a) is that it involves {\em no}
additional parameters beyond those used to obtain a chiral Equation of
State fit.  Either Eq. (14a) fits the spectroscopy data or the
hypothesis of logarithmic triviality is wrong.  Eq. (14a) simplifies
at the critical point $\beta=\beta_c$ to,
\begin{equation}
R(\beta_c,m)=(3+\ln^{-1}|\sigma^{-1}|)^{-1}
\end{equation}
The ``three'' in Eq. (15) occurs because
$\delta=3$ in mean field theory.  So, in this model as opposed to a
nontrivial fixed point theory, $R$ is not quite constant at the
critical point and it approaches its chiral limit of ${1\over 3}$ from
below.  In Fig.~9 we plot Eq. (14) and see that it differs
qualitatively from the spectroscopy data.  For example, the
spectroscopy data for $R$ falls as $m$ decreases at $\beta=.190$ while
the logarithmically-improved $0(2)$ sigma model predicts that it should
rise.  The data changes from concave downward to concave upward for
$\beta$ between  .20 and .21 indicating that the critical point is
in the interval while the logarithmically-improved $0(2)$ model curves
change their character for $\beta$ between .18 and .19 because
$\beta_c=.186$ is incorporated in the chiral Equation of State
Eq. (13).

\section{The Universality Class of Noncompact\protect\\Lattice QED}

We noted in Sec.~3 above that the critical indices of our power-law
fits to the $N_f=4$ chiral condensate data were essentially the same as
those obtained previously for the $N_f=2$ model.  Thus, the two models
should be in the same universality class and should have identical
universal functions~$f$ (up to a choice of scale) and Equation of
State.  We, therefore, plot in Fig.~9 the Equation of State (6) for
both the $N_f=2$ data of Ref.~[2] and the $N_f=4$ data discussed here.
A linear fit and universality is quite compelling.  Note that the
$N_f=2$ data was obtained at different couplings, different bare
fermion masses, by a different algorithm, on a lattice of different
size, on a different computer.

This intriguing result begs for further justification.  Larger scale
more accurate mass spectrum and chiral condensate measurements are
imperative.  Will these results persist on larger lattices or will
triviality eventually be found?  Can we implicate the monopoles more
directly in the physics of the chiral phase transition?  How can it be
that the critical indices are essentially independent of the number of
flavors?  How can fermion screening shift the critical couplings as a
function of flavor but not effect the dynamics of the critical point?
This is just a sampling of the array of questions the results of
Ref.~[2] and this paper produce.

We note with interest that the methods of Ref.~[9] have predicted
critical couplings in excellent agreement with ours.

\newpage

\section*{Acknowledgement} The simulation done here used the CRAY
facilities of NERSC and PSC as well as the now defunct ST-100 of
Argonne National Laboratory.  J. B. Kogut is supported in part by
NSF-PHY97-00148.

\section*{References}
\begin{enumerate}

\item E. Dagatto, A. Kocic and J. B. Kogut, {\em Phys.\ Rev.\ Lett.}.
{\bf 60}, 772 (1988); {\bf 61}, 2416 (1988).

\item S. J. Hands, R. L. Renken, A. Kocic, J. B. Kogut, D. K.
Sinclair and K.C. Wang, {\em Phys.\ Lett.} {\bf B261}, 294 (1991)
ILL-(TH)-92-\#16, Aug., 1992.

\item S. J. Hands, A. Kocic, and J.B. Kogut, ILL-(TH)-92-7 (to appear
in {\em Phys.\ Lett.\ B})

\item S. J. Hands, A. Kocic, and J. B. Kogut, {\em Nuc.\ Phys.} {\bf
B357}, 467 (1991).

\item M. Gockeler, R. Horsley, P. Rakow, G. Schierholz and R. Sommer,
{\em Nuc.\ Phys.} {\bf B371}, 713 (1992)

\item A. Kocic, J. B. Kogut and M.-P. Lombardo, ILL-(TH)-92-18, Aug.,
1992.

\item S. J. Hands and R. Wensley, {\em Phys.\ Rev.\ Lett.} {\bf 63},
2169 (1989).

\item A. Kocic, J. B. Kogut, M.-P. Lombardo and K. C. Wang,
ILL-(TH)-92-12, CERN-TH. 6542/92, June, 1992.

\item A. Azcoiti, G. Di~Carlo and A. F. Grillo, DFTUZ.91/34 and
references contained therein.

\item S.J. Hands, A. Kocic, E. Dagatto and J.B. Kogut, {Nuc.\ Phys.}
{\bf B347}, 217 (1990).

\end{enumerate}

\newpage

\begin{table}[hbt]
\begin{center}
\caption{Monopole Susceptibility $\chi$ on a $10^4$ Lattice.}
\scriptsize
\vspace{12pt}
\begin{tabular}{lllllllll}
$\beta\backslash m$&.005&.01&.02&.03&.04&.05&.06&.07\\
.22&32.1(7)&32.9(6)&---&36.5(7)&37.3(1.2)&41.2(9)&43.3(9)&45.7(8)\\
.215&39.0(8)&41.5(1.4)&---&47.5(2.6)&49.9(1.5)&48.5(1.3)&49.6(2.3)&50.8(1.6)
\\
.21&48.2(1.6)&49.7(1.4)&---&48.3(2.5)&45.8(2.2)&48.7(2.0)&45.3(1.9)&44.6(2.3
)\\
.205&52.9(3.0)&49.1(1.8)&---&42.1(3.3)&34.6(1.9)&35.9(2.1)&27.2(1.9)&30.3(1.
9)\\
.20&41.4(2.6)&34.9(2.6)&---&28.1(3.2)&17.8(1.5)&16.4(1.6)&11.6(7)&10.3(7)\\
.195&17.1(1.6)&16.0(1.5)&---&9.2(6)&7.6(4)&6.6(3)&5.9(2)&5.3(1)\\
.19&8.4(6)&7.0(4)&---&4.9(3)&4.3(3)&3.5(1)&3.4(1)&3.2(1)\\
.185&---&---&3.2(1)&2.8(1)&2.6(1)&2.3(1)&2.2(1)&2.0(1)\\
\end{tabular}
\end{center}
\end{table}

\begin{table}[hbt]
\begin{center}
\caption{Monopole Percolation Order Parameter on a $10^4$ Lattice.}
\scriptsize
\vspace{12pt}
\begin{tabular}{lllllllll}
$\beta\backslash m$  &.005  &.01  &.02  &.03  &.04  &.05  &.06  &.07\\
.22  &.070(2)  &.077(2)  &---  &.083(2)  &.104(9)  &.110(5)  &.117(5)
&.130(5)\\
.215  &.104(3)  &.104(3)  &---  &.131(4)  &.147(6)  &.176(5)  &.186(8)
&.208(6)\\
.21  &.154(4)  &.159(6)  &---  &.223(9)  &.259(11)  &.283(8)  &.298(8)
&.336(7)\\
.205  &.225(8)  &.270(8)  &---  &.347(9)  &.396(7)  &.417(7)  &.457(8)
&.480(4)\\
.20  &.373(9)  &.407(8)  &---  &.478(7)  &.533(5)  &.565(6)  &.589(4)
&.605(4)\\
.195  &.523(7)  &.548(9)  &---  &.619(5)  &.642(4)  &.666(6)  &.681(3)
&.695(3)\\
.19  &.639(4)  &.659(4)  &---  &.708(3)  &.723(3)  &.744(2)  &.752(1)
&.760(2)\\
.185  &---  &---  &.759(2)  &.776(2)  &.788(2)  &.798(2)  &.807(2)
&.816(1)\\
\end{tabular}
\end{center}
\end{table}

\begin{table}[htb]
\begin{center}
\caption{Monopole Concentration on a $10^4$ Lattice.}
\scriptsize
\vspace{12pt}
\begin{tabular}{lllllllll}
$\beta\backslash m$  &.005  &.01  &.02  &.03  &.04  &.05  &.06  &.07\\
.22  &.1163(2)  &.1171(2)  &---  &.1194(4)  &.1207(3)  &.1223(3)  &.1237(2)
 &.1249(2)\\
.215  &.1224(2)  &.1229(3)  &---  &.1260(3)  &.1275(3)  &.1289(3)
&.1301(3)  &.1315(2)\\
.21  &.1287(2)  &.1292(3)  &---  &.1331(4)  &.1349(4)  &.1361(2)  &.1371(3)
 &.1388(3)\\
.205  &.1351(3)  &.1360(2)  &---  &.1401(3)  &.1422(3)  &.1437(3)
&.1452(3)  &.1469(2)\\
.20  &.1422(3)  &.1432(4)  &---  &.1472(3)  &.1501(2)  &.1518(3)  &.1530(3)
 &.1548(3)\\
.195  &.1496(4)  &.1512(3)  &---  &.1560(3)  &.1582(3)  &.1602(4)
&.1617(3)  &.1636(3)\\
.19  &.1584(2)  &.1599(4)  &---  &.1653(3)  &.1670(4)  &.1695(3)  &.1708(2)
 &.1723(3)\\
.185  &---  &---  &.1723(3)  &.1748(3)  &.1770(3)  &.1784(3)  &.1802(2)
&.1820(3)\\
\end{tabular}
\end{center}
\end{table}

\begin{table}
\begin{center}
\caption{$16^4$ data at $m=.03$ for the plaquette $S_o$, the chiral
condensate $\bra\bar\psi\psi\ket$, the monopole susceptibility $\chi$,
the monopole order parameter $M$ and the monopole concentration
$\bra\l\ket/N_\l$}
\scriptsize
\vspace{12pt}
\begin{tabular}{llllll}
$\beta$  &$S_0$  &${\bra\bar\psi\psi\ket}$  &$\chi$  &$M$
&${\bra\ell\ket/N\ell}$\\
.22  &.9751(5)  &.1625(5)  &62.7(8)  &.0287(7)  &.1199(1)\\
.215  &.9950(3)  &.1746(5)  &97.4(1.2)  &.0550(15)  &.1257(1)\\
.21  &1.0181(4)  &.1899(6)  &146.0(4.6)  &.1452(41)  &.1326(1)\\
.205  &1.0442(4)  &.2076(6)  &61.6(4.2)  &.3533(45) & .1405(2)\\
.20  &1.0710(5)  &.2259(5)  &17.2(5)  &.5196(25)  &.1486(2)\\
.195  &1.0992(5)  &.2450(6)  &8.3(2)  &.6294(21)  &.1568(2)\\
.19  &1.1288(6) & .2658(8)  &4.7(1)  &.7112(12)  &.1655(2)\\
\end{tabular}
\end{center}
\end{table}

\begin{table}
\begin{center}
\caption{Plaquette averages on a $10^4$ Lattice}
\scriptsize
\vspace{12pt}
\begin{tabular}{lllllllll}
$\beta\backslash m$  &.005  &.01  &.02  &.03  &.04  &.05  &.06  &.07\\
.22  &.9638(6)  &.9661(5)  &.9679(4)  &.9735(10)  &.9779(10)  &.9829(8)
&.9869(6)  &.9906(8)\\
.215  &.9842(6)  &.9851(7)  &.9904(6)  &.9952(8)  &1.0003(7)  &1.0046(8)
&1.0087(8)  &1.0136(6)\\
.21  &1.0060(5)  &1.0073(6)  &1.0135(6)  &1.0197(9)  &1.0247(10)
&1.0287(5)  &1.0320(6)  &1.0374(7)\\
.205  &1.0263(5)  &1.0301(7)  &1.0361(60  &1.0430(8)  &1.0485(7)
&1.0536(7)  &1.0585(6)  &1.0636(6)\\
.20  &1.0505(7)  &1.0541(9)  &1.0621(8)  &1.0671(8)  &1.0766(7)  &1.0810(8)
 &1.0858(7)  &1.0913(6)\\
.195  &1.0761(9)  &1.0805(9)  &1.0896(9)  &1.0969(9)  &1.1036(10)
&1.1098(11)  &1.1152(7)  &1.1216(10)\\
.19  &1.1051(6)  &1.1113(12)  &1.1118(7)  &1.1282(9)  &1.1349(8)
&1.1424(9)  &1.1469(6)  &1.1515(10)\\
.185  &---  &---  &1.1526(8)  &1.1612(7)  &1.1681(10)  &1.1729(9)
&1.1790(7)  &1.1848(8)\\
\end{tabular}
\end{center}
\end{table}

\begin{table}
\begin{center}
\caption{Chiral condensate on a $10^4$ Lattice}
\scriptsize
\vspace{12pt}
\begin{tabular}{lllllllll}
$\beta\backslash m$  &.005  	&.01  &.02  &.03  &.04  &.05  &.06  &.07\\
.22  	&.0325(4)  	&.0658(6)  &.1144(7)    &.1588(16)  &.1921(15)
&.2173(10)  &.2387(9)  &.2572(8)\\
.215  &.0370(5)  	&.0729(7)  &.1292(8)   &.1737(10)  &.2054(11)  &.2303(12)
 &2519(9)  &.2695(8)\\
.21  	&.0491(12)  &.0883(8)  &.1466(10   &.1901(9)   &.2213(9)   &.2434(11)
 &.2640(9)  &.2823(8)\\
.205  &.0580(13)  &.1059(10) &.1623(12) &.2058(13)  &.2325(13) &.2575(9)
&.2781(8)  &.2969(8)\\
.20  	&.0754(17)  &.1244(18) &.1875(13)  &.2214(13)  &.2523(9)  &.2754(8)
&.2938(6)  &.3104(8)\\
.195  &.0986(21)  &.1533(21) & .2065(14)  &.2420(9)  &.2696(12) & .2905(13)
 &.3105(9)  &.3271(8)\\
.19  	&.1335(23)  &.1820(21) &.2282(16)  &.2650(12)  &.2893(14) &.3111(15)
&.3257(8)  &.3400(1)\\
.185  &---  						&---  					&.2579(10)  &.2887(10)  &.3095(12)  &.3261(9)
&.3410(9)  &.3554(8)\\
\end{tabular}
\end{center}
\end{table}

\begin{table}[tp]
\begin{center}
\caption{Finite size effects in $\bra\bar\psi\psi\ket$ on $10^4$ and
$12^4$ Lattices}
\scriptsize
\vspace{12pt}
\begin{tabular}{lllll||llll}
\multicolumn{5}{c||}{$10^4$ Lattice}&\multicolumn{4}{c}{$12^4$ Lattice}\\
$m\backslash \beta$ &.19       &.20       &.21      &.22       &.19
&.20      &.21      &.22\\
.04                 &.2893(14) &.2523(9)  &.2213(9) &.1921(15) &.2892(6)
&.2514(5) &.2197(4) &.1917(4)\\
.02                 &.2282(16) &.1875(13) &.1466(10 &.1144(7)  &.2340(7)
&.1891(6) &.1550(6) &.1213(4)\\
.01                 &.1820(21) &.1244(18) &.0883(8) &.0658(6)  &---
&1322(10) &.0917(6) &---\\
\end{tabular}
\end{center}
\end{table}
\vfill
\eject

\clearpage
\section*{Figure Captions}
\begin{enumerate}
\def\theenumi{Figure~\arabic{enumi}}

\item{Monopole Susceptibility on $10^4$ (circles) and $16^4$
(triangles) lattices at $m=.03$}

\item{Monopole Order Parameter $M$ (circles) and
Concentration (triangles) on a $16^4$ Lattice at $m=.03$}

\item{$\ln{\bra\bar\psi\psi\ket}$ vs.\ $\ln(m)$ at Various
Couplings ($\beta=.185$ (dark circle), .190 (dark square), .195
(inverted dark triangle), .200 (dark triangle), .205 (cross), .210
(triangle), .215 (inverted triangle), .220 (square)) on a $10^4$
Lattice}

\item{$-1/\ln(m)$ vs.\
$-1/\ln{\bra\bar\psi\psi\ket}$ on a $10^4$ Lattice}

\item{Chiral Equation of State Eq. (6) for $10^4$
$\bra\bar\psi\psi\ket$ data $m=.03-.07$, $\beta=.22$--.185}

\item{$R$($\beta,m$) vs.\ $m$ for Various $\beta$ for the $12^4$
Spectroscopy Data of Ref.~[5]}

\item{$R(\beta,m)$ vs.\ $m$ for Various $\beta$ of Fig.~6 From the Fixed
Point Equation of State (EOS) of Eq. (6) and (9)}

\item{Chiral Equation of State Fit of the $\beta=.22$--.185,
$m=.03$--.07 $10^4$ Data Using the Logarithmically-improved O(2) Model
of Ref.~[5]}

\item{$R(\beta,m)$ vs.\ $m$ for Various $\beta$ of Fig.~6 From the
Logarithmically-improved O(2) Sigma Model of Ref.~[5]}

\item{Fixed Point Equation of State for $N_f=2$ Data of Ref.~[2] and
$N_f=4$ Data of Table 6}
\end{enumerate}

\end{document}